\documentclass[aps,prl,reprint]{revtex4-2}

\usepackage{graphicx}
\usepackage{dcolumn}
\usepackage{bm}
\usepackage{amsmath}
\usepackage{xcolor}
\usepackage{ulem}

\begin{document}

\preprint{APS/123-QED}

\title{Capillary jet in a slab of foam}

\author{Théophile Gaichies}
\author{Bryan Giraud}%
\author{Emmanuelle Rio}%
\affiliation{Université Paris-Saclay, CNRS, Laboratoire de Physique des Solides, 91405, Orsay, France
}%

\author{Anniina Salonen}%
\affiliation{Soft Matter Sciences and Engineering, ESPCI Paris, PSL University, CNRS, Sorbonne Université, 75005, Paris}
\altaffiliation[Also at ]{Université Paris-Saclay, CNRS, Laboratoire de Physique des Solides, 91405, Orsay, France}

\author{Arnaud Antkowiak}
\affiliation{
Institut Jean le Rond $\partial$'Alembert, Sorbonne Université, CNRS, F-75005 Paris, France
}%

\date{\today}

\begin{abstract}

A capillary jet plunging into a quasi-2D slab of monodisperse foam of the same solution is studied experimentally. We show that the jet can have a drastic impact on the foam. At small speeds it inflates the channels separating the bubbles. At intermediate speeds, impact with channels creates very small bubbles. At higher speeds, the jet  thins the foam
films, which can result in bubble rupture. We study the mechanisms behind these processes using an elementary foam formed by three soap films in contact through a Plateau border. We obtain scaling laws, which transfer quantitatively from the elementary foam to the 2D foam.

\end{abstract}

\maketitle


In breaking waves, weirs or wastewater plants, liquid can entrain air as it impacts the liquid reservoir \cite{kiger2012air,erinin2023effects,liu2003effects}. If the solution contains surfactant, bubbles can form and accumulate at the surface. The liquid will then impact a foam. The same configuration is found in the Ross Miles test, where the quantity of foam produced by drops hitting a foaming solution is measured to characterize its foaming properties \cite{cheah2005foaming,ross1941apparatus}. Interactions between foam and a flowing fluid have also been studied in the context of oil extraction. The fluid was either air or pure water, and it generally lead to foam rupture \cite{park1994viscous,mensire2014coalescence,del2013blast,salem2013response,arif2010speed,salkin2016generating}.
Anti-shock properties of foam motivated studies on the impact of drops or solid objects in soap films or foams\cite{gilet2009fluid,goff2014shooting,goff2013supershear,monloubou2015influence,monloubou2016blast,cohen2017raufaste,tani2021dynamics}. Although they mimic most natural situations cited above, studies with jets of foaming solution were confined to an inclined jet impacting a single film \cite{kirstetter2012}. 

In this letter, we study experimentally the interaction between a capillary jet and a layer of foam sandwiched between two glass plates (2D foam).
The cell consists of two glass plates (17.8 mm x 12.7 mm x 1 mm) glued together with a chemical resistant epoxy adhesive (Araldite 2014). A stainless steel nozzle is fixed at the bottom of the cell. We vary the spacing $d$ between the two glass plates between 2 and 6 mm with 3D printed wedges. The foam is produced by bubbling
in a 5 wt$\%$ Fairy solution, with a pressure controller. This allows us to produce monodisperse foams with a mean bubble radius $R_b$ between 1.2 and 4.3 mm. A jet of the foaming solution is generated by imposing a pressure in a liquid reservoir with a pressure controller (Elve-
flow OB1 Mk4). This reservoir is connected to nozzles of varying sizes to study laminar jets with radii $R$ between 0.075 and 0.73 mm and speeds $V$ between 0.4 and 6 $\mathrm{m.s^{-1}}$. To perform an experiment, we bubble into the solution to fill the cell to the brim with foam. The jet then plunges vertically at the top of the cell while the system is recorded with a camera (Basler a2A1920-160umPRO) and a fixed focal length objective (C23-3520-2M f35mm) during at least ten seconds at twenty frames per second. 
\begin{figure}[h!]
    \centering
    \includegraphics[width=0.46\textwidth]{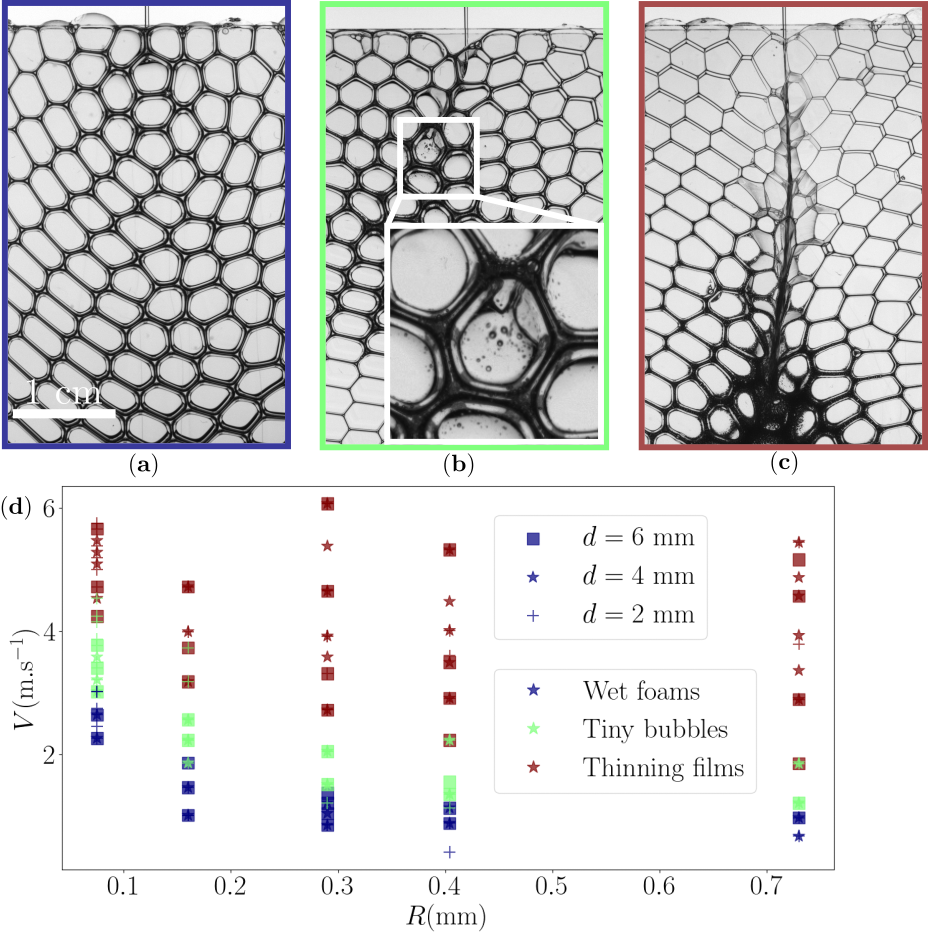}
    \caption{(a),(b) and (c) : Jet of radius 0.16 mm moving respectively at  $V = 1\  \mathrm{m.s^{-1}}$, $V = 2.2\  \mathrm{m.s^{-1}}$, $V = 4.7 \ \mathrm{m.s^{-1}}$, in a monodisperse foam  with $R_b = 1.6 \ \mathrm{mm}$, in a cell with a spacing $d=2 \mathrm{mm}$. (d) Phase diagram showing the classification of experiments in three regimes for different speeds and radii.}
    \label{fig1}
\end{figure}

We can classify all the experiments into three regimes: "Wet foams", "Tiny bubbles" and "Thinning films". Photographs in each regime can be found in Fig.\ref{fig1}. At low speed the jet simply inflates the liquid channels, called Plateau borders. It wets the foam, as evidenced by  its darkening in  Fig.\ref{fig1}(a). At intermediate speed the jet produces tiny bubbles within the foam channels (Fig.\ref{fig1}(b)). These tiny bubbles have a radius much smaller than $R_b$ and although visible in Fig.\ref{fig1}(b), most of them accumulate in foam channels. In the third regime, the Plateau borders and the films are drained in the vicinity of the jet as assessed by the lighter color of the channels (Fig.\ref{fig1}(c))\footnote{See video examples in the supplementary materials\cite{supp}}. The thinning of the films can ultimately lead to bubble rupture. Note that tiny bubbles are still produced in this regime, and they accumulate at the bottom of the foam. We sum up the results of the experiments in a phase diagram in Fig.\ref{fig1}(d). It appears that neither the bubble size nor the spacing $d$ plays a major role in the effect of the jet on the foam. The phase diagram is thus controlled by $V$ and $R$, with reduced transition speeds  with increasing $R$. 

\begin{figure}[h!]
    \centering
    \includegraphics[width=0.46\textwidth]{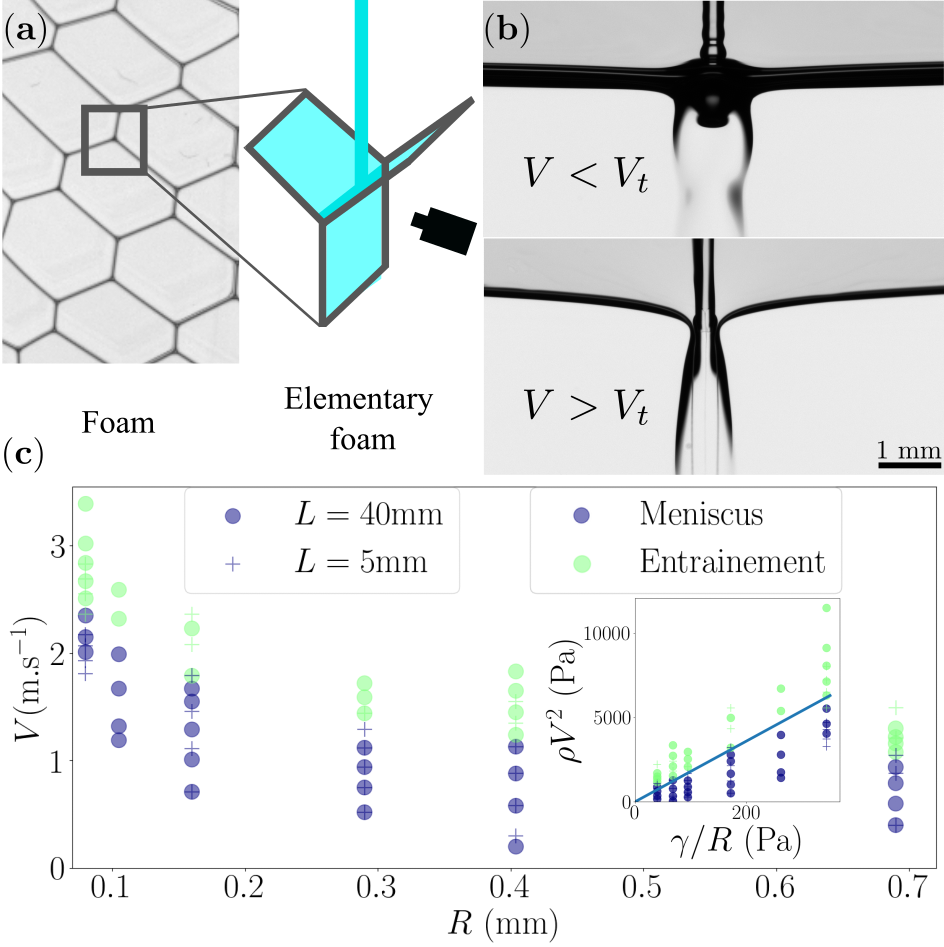}
    \caption{(a) Photograph of a foam and sketch of the elementary foam on which we study the interaction between a jet and a Plateau border. (b) Experimental pictures of 0.16 mm jets impacting a 40 mm wide Plateau border at $V = 1.1 \ \mathrm{m.s^{-1}}$, for the top picture and $V = 2.2 \ \mathrm{m.s^{-1}}$ for the bottom one. At the connection between the jet and the plateau border, we observe respectively a meniscus toward the jet and an inverse meniscus that allows the presence of small air jets. (c) Phase diagram of jets impacting Plateau border, with an inset displaying the same data plotted according to the scaling.}
    \label{fig2}
\end{figure}

To study the transitions between these regimes, we create a simplified experiment. It  consists of three soap films of equal dimensions, connected at 120° angles through a Plateau border(see Fig.\ref{fig2}(a)). This system can be seen as an elementary unit of foam, and it allows us to control the orientation between the films and the jet. The elementary foam is formed for each experiment by dipping the frame in our foaming solution. To reproduce the transition from Wet foams to Tiny bubbles, the vertical jet directly  impacts the Plateau border, as shown in Fig.~\ref{fig2}(a). We perform these experiments with the same jet radii as in the 2D foam experiment, with two different Plateau border lengths $L$. We can observe in Fig.\ref{fig2}(b), that at low speeds, the meniscus between the jet and the Plateau border turns up. Stationary capillary waves can be observed at the base of the jet, which are due to a pressure mismatch between the jet and the Plateau border \cite{gaichies2024}. Above a critical speed, we observe an inversion of the meniscus that allows the jet to entrain air threads in the film below. These thin threads of air destabilise into multiple bubbles. The radius of the air threads varies with the position of the jet relative to the Plateau border, but remains significantly smaller than $R$. We then classify our experiments into two categories depending on the curvature of the meniscus. As seen in Fig.\ref{fig2}(c), the results do not depend on $L$. To model this transition, we compare the dynamical pressure of the jet $\rho V^2$ (where $\rho = 10^3 \mathrm{kg.m^{-3}}$ is the density of the foaming solution), which pushes down on the Plateau border, with capillary forces that resist this deformation. The typical curvature of the deformation of the Plateau border is $1/R$, so the resisting capillary pressure scales as $\gamma/R$, where $\gamma = 0.028 \ \mathrm{N.m^{-1}}$ is the surface tension of our solution. As seen in the inset of Fig.\ref{fig2}(c), this scaling captures the transition between the two regimes, with a prefactor of $18$.

\begin{figure}[h!]
    \centering
    \includegraphics[width=0.46\textwidth]{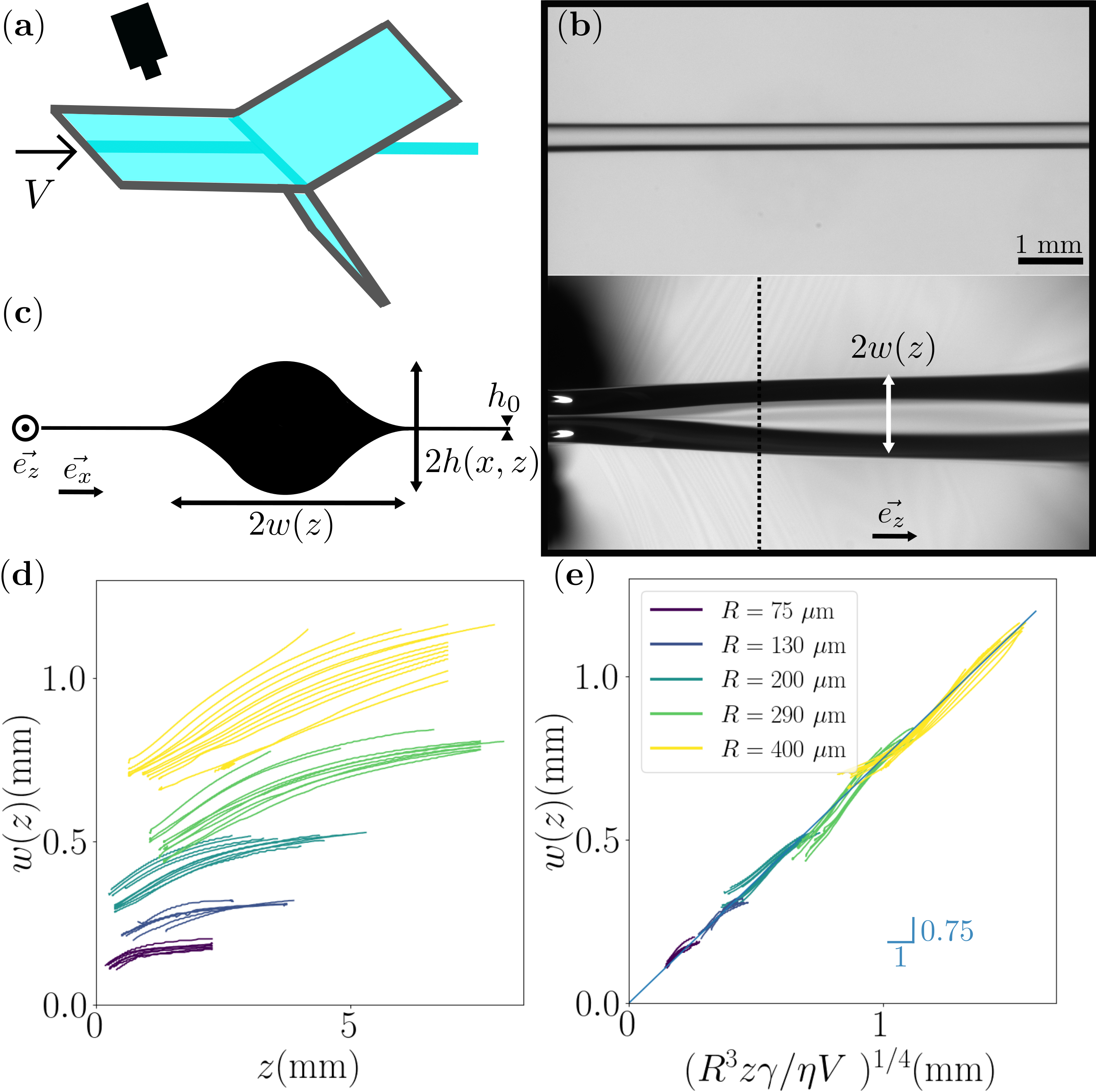}
    \caption{(a) Sketch of the experimental setup, where we observe the evolution of a jet in a film in reflection. (b) Experimental photograph of a jet with $R=0.2 \ \mathrm{mm}$ at $V= 5.2 \ \mathrm{m.s^{-1}}$ in air (top), and in a soap film (bottom). (c) Sketch of the cross-section along the dotted line on (b). (d) Extracted profiles of the jet. (e) Extracted profiles of the jet plotted against the scaling of Eq. (1).}
    \label{fig3}
\end{figure}

To study the transition between the regime of Tiny bubbles and Thinning films we inject directly into the film through a small opening in the frame. The jet flows through the film of the elementary foam and exits through a Plateau border, as depicted in Fig.\ref{fig3}(a). We observe the film in reflection. The jet radius is impacted by the presence of the soap film. In air (Fig.\ref{fig3}(b)(top)), the jet is straight, and its radius stays constant in the field of view of the imaging system. Inside a film (Fig.\ref{fig3}(b)(bottom)), the width of the jet $2w$ increases as it travels through the film. This behavior is stationary in time, as long as a film is present. As the speed of the jet increases, this widening is reduced. To quantify these observations, we extract the profile $w(z)$ from experimental images, through image processing using scikit-image \cite{van2014scikit}. The experimental profiles for various speeds and radii are plotted in Fig.\ref{fig3}(d). To model this evolution, we use the notation sketched in Fig.\ref{fig3}(c), which represents the jet cross section along the dotted line in Fig.\ref{fig3}(b)(bottom). We assume that the evolutions in the $x$ and z directions are decorrelated so we have an equivalence between time and space $t=z/V$. The evolution of the jet along the $z$ direction is thus the same as a 2D drop flattening in time. The drive for such spreading is the drop overpressure $\gamma/R$ compared to the flat film. We describe the evolution of the jet's height $h(x,t)$ with the lubrication equation : 
\begin{equation*}
    \frac{\partial h}{\partial t} = - \frac{\gamma}{3\eta} \frac{\partial }{\partial x}(h^3\frac{\partial^3 h}{\partial x^3}),
\end{equation*}
where $\eta = 1 \ \mathrm{mPa.s^{-1}}$ is the viscosity of our solution. We obtain a scaling by taking $h \sim R$, and $x \sim w$ as proposed by McGraw \textit{et al.} in a similar problem \cite{mcgraw2012self}. This yields, using $t=z/V$ : 
\begin{equation}
    w(z)\sim (\frac{R^3z\gamma}{\eta V})^{1/4},
\end{equation}
in good agreement with the measured profiles $w(z)$, with a prefactor of $0.75$ (Fig.\ref{fig3}(e)). 
\begin{figure}[h!]
    \centering
    \includegraphics[width=0.5\textwidth]{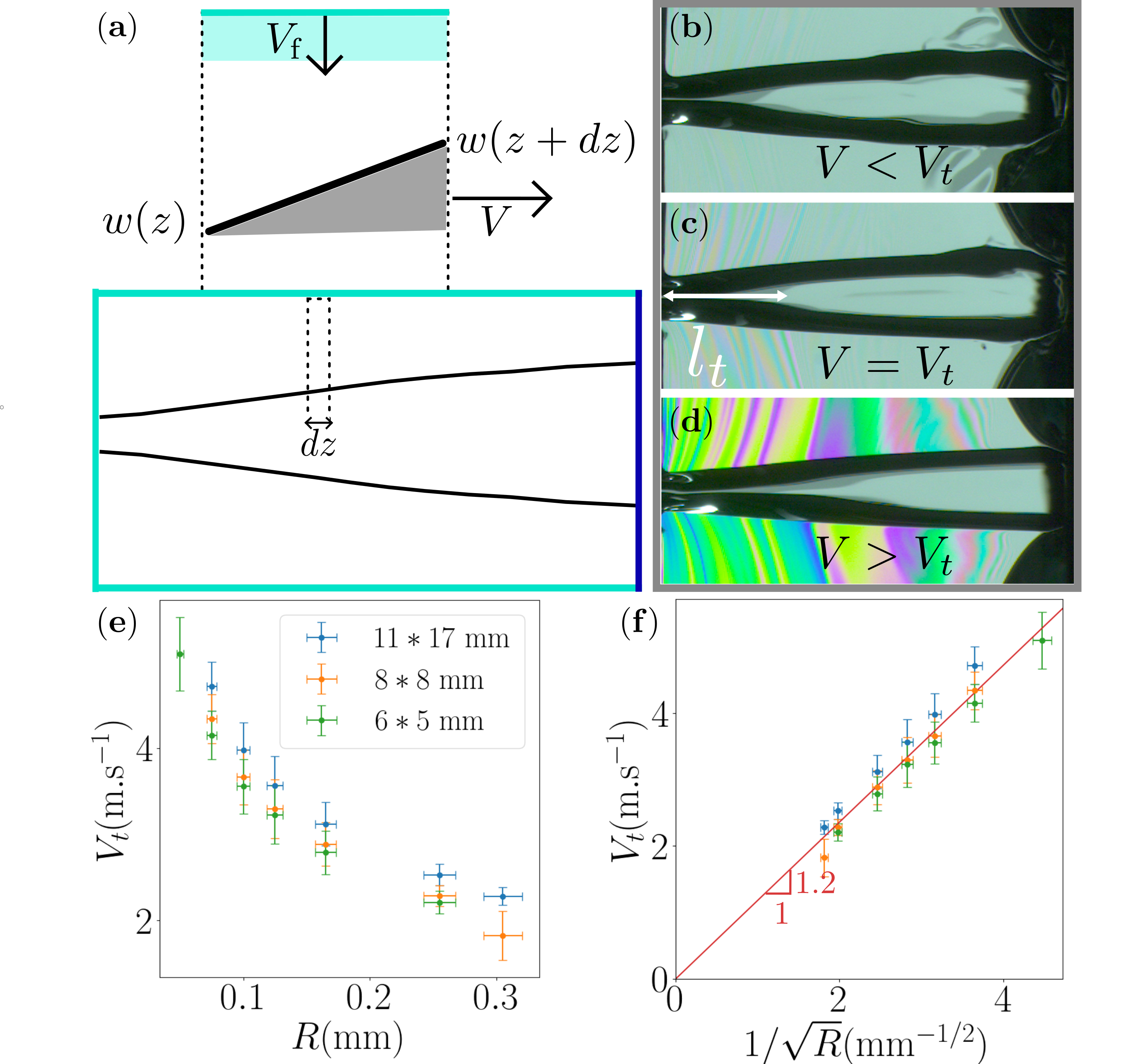}
    \caption{(a) Sketch of surface fluxes, extracted from the meniscus, or caused by the spreading of the jet. (b) and (c) Photograph of a jet of $R=0.4 \ \mathrm{mm}$ in a soap film at (b) $V= 2 \ \mathrm{m.s^{-1}}$, below the transition (c) $V= 2.5 \ \mathrm{m.s^{-1}}$,at the transition, where colors appear and (d) $V= 3.9 \ \mathrm{m.s^{-1}}$, above the transition, where colors invade the film.(e) Speed at which a stable colored band appears for different jet radii in elementary foams. (f) The same data plotted according to the scaling of Eq. (2).}
    \label{fig4}
\end{figure}

If we now focus on the surrounding soap film, we can see that it experiences a transition similar to the one observed in 2D foams, as shown in Fig.\ref{fig4}. Below the transition speed, the film stays thick and colorless, and has an increased lifetime compared to a film in the absence of a jet (Fig.\ref{fig4}(b)). At a given speed, a stationary colored band of width $l_t$ appears in the film close to the injection point of the jet (left in  Fig.\ref{fig4}(c)). The width of this colored band grows with the speed of the jet (Fig.\ref{fig4}(d)). The colors indicate that the film thickness becomes submicrometric. To characterize this transition, we measure the velocity $V_t$ where this band appears, as a function of $R$ in elementary foams of different sizes. The results of these experiments are plotted in Fig.\ref{fig4}(e). $V_t$ is mostly independent of the size of the film, and controlled by $R$. To model this transition, we consider a surface of width $dz$ as depicted in Fig.\ref{fig4}(a). As $w$ increases with $z$, a part of the film surface is entrained by the jet. The rate of entrainment is proportional to $V$, as is the case in the entrainment by a 3D rising plume in convective flows \cite{morton1956turbulent}. This surface flux can thus be written $F\sim V\mathrm{d}z\partial w /\partial z $. The missing surface can be replaced either by fresh film, or by stretching an existing part of the soap film. Fresh surface must be extracted from a meniscus. This extraction results from a difference in surface tension $\Delta \gamma$ between the film and the meniscus, as explained by Frankel \cite{frankel}. $\Delta \gamma = E \text{d}A/A$ is positive, because of surface extraction. As long as $\text{d}A/A << 1$, the surface can be considered incompressible. Thus, $F$ must be entirely compensated by the flux of fresh film, so that $F=F_f \sim V_f dz$, where $V_f=\Delta \gamma^{3/2}/\eta\gamma^{1/2}$ is the Frankel velocity \cite{frankel}. However, when $dA/A \sim 1 $, the incompressibility assumption is no longer valid, as the film starts to stretch and thin. The transition between compressible and incompressible interfaces, to which we attribute the apparition of the color band, thus corresponds to $\Delta \gamma \sim E$. As the colored band stretches up to $l_t$, we evaluate $\partial w /\partial z$ at $z=l_t$, which gives the transition velocity 
\begin{equation}
    V_t \sim \frac{l_t E^2}{R\gamma \eta}.
\end{equation}
To compare this scaling with experiments, we have to explicit $l_t$. Experimentally (see supplementary materials \cite{supp}), we obtain the empirical fit $l_t \sim 0.15 R^{1/2}$. In Fig.\ref{fig4}(e), we plot $V_t$ against $1/\sqrt{R}$ to test this scaling. The data align along a line of slope $1.2$, which corresponds to $E=2.6 \ \mathrm{mN.m^{-1}}$, close to values from literature. The Gibbs elasticity is mostly fixed by surfactant concentration and film thickness \cite{mysels61,vandentempel65}. For micrometric thicknesses at a concentration of a few times the critical micellar concentration, an elasticity of a few mN/m is then reasonable \cite{poryles22,prins67}. 
\begin{figure}[h!]
    \centering
    \includegraphics[width=0.46\textwidth]{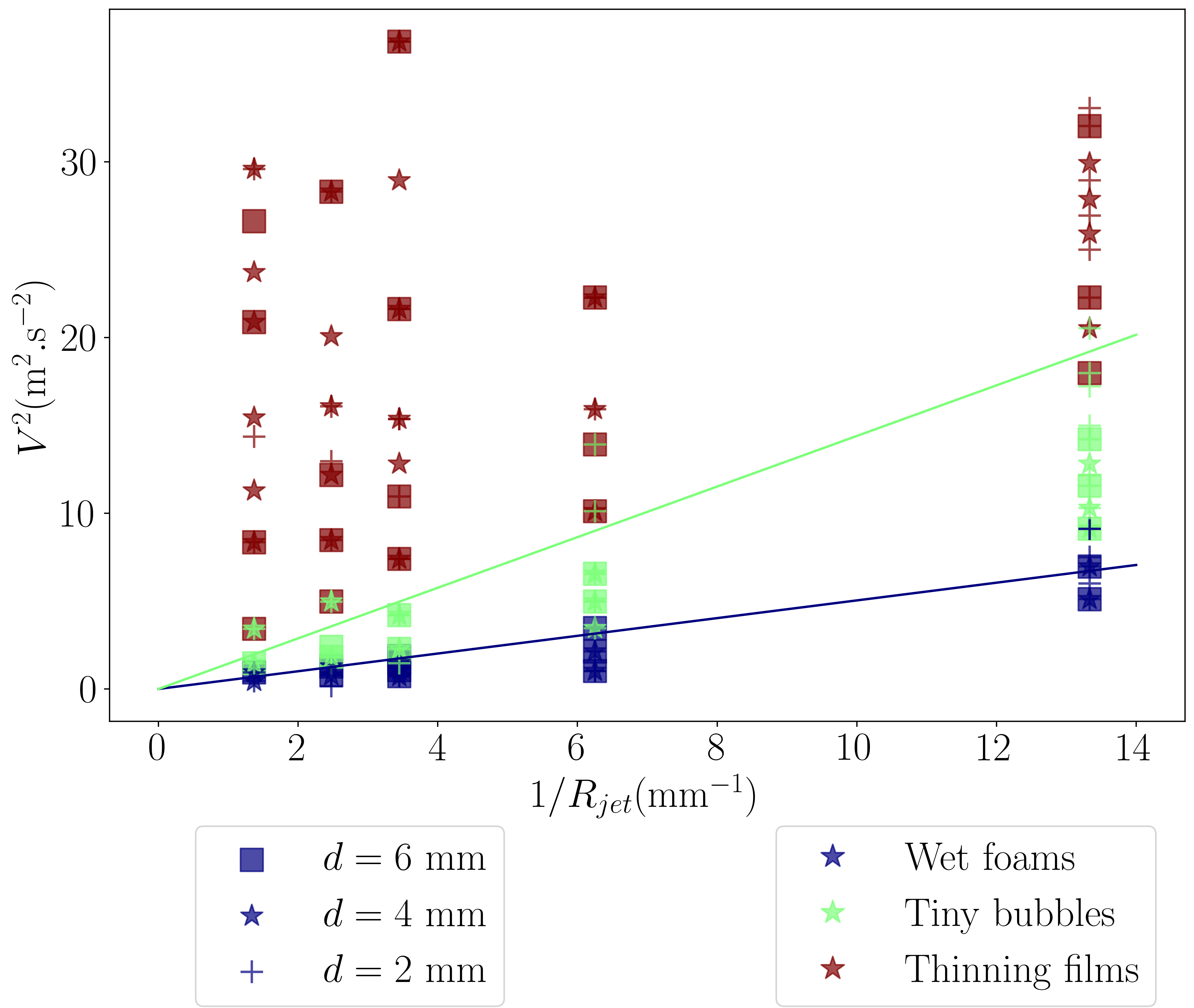}
    \caption{ Phase diagram with two lines that correspond to the scaling found in elementary foam i.e $V^2 = 0.0005/R$ for the blue line and $V^2 = 0.0014/R$ for the green line in S.I units.  }
    \label{fig5}
\end{figure}

We have captured the transitions in the elementary foam, and we can now return to the 2D foam. In Fig.\ref{fig5}, we plot the scaling laws from Fig.\ref{fig2}(c) and Fig.\ref{fig4}(e) against the data of the 2D foam experiments. We obtain a quantitative agreement between the experiments at the foam and film scales. In each case, the scaling corresponds to an effective Weber number $\mathrm{We}=\rho V^2R/\gamma$, $\mathrm{We}=18$ for Tiny bubbles and $\mathrm{We}=50$ for Thinning films in our system.

In conclusion, we have identified three regimes of interaction between a capillary jet and a 2D foam. The wet foam regime resembles forced drainage, in which the local liquid fraction increases. In a second regime, the presence of tiny bubbles impacts drastically the bubble size distribution in the foam, which has implications on its stability and mechanical properties. In the thinning films regime, the jet can even destroy the upper bubbles. The mechanisms and associated scaling laws were deduced from experiments at the film scale, we thus expect the transitions to hold in 3D foams. These interactions should be considered in applications where a jet impacts continuously a bubbling solution. 


\begin{thebibliography}{30}%
\makeatletter
\providecommand \@ifxundefined [1]{%
 \@ifx{#1\undefined}
}%
\providecommand \@ifnum [1]{%
 \ifnum #1\expandafter \@firstoftwo
 \else \expandafter \@secondoftwo
 \fi
}%
\providecommand \@ifx [1]{%
 \ifx #1\expandafter \@firstoftwo
 \else \expandafter \@secondoftwo
 \fi
}%
\providecommand \natexlab [1]{#1}%
\providecommand \enquote  [1]{``#1''}%
\providecommand \bibnamefont  [1]{#1}%
\providecommand \bibfnamefont [1]{#1}%
\providecommand \citenamefont [1]{#1}%
\providecommand \href@noop [0]{\@secondoftwo}%
\providecommand \href [0]{\begingroup \@sanitize@url \@href}%
\providecommand \@href[1]{\@@startlink{#1}\@@href}%
\providecommand \@@href[1]{\endgroup#1\@@endlink}%
\providecommand \@sanitize@url [0]{\catcode `\\12\catcode `\$12\catcode
  `\&12\catcode `\#12\catcode `\^12\catcode `\_12\catcode `\%12\relax}%
\providecommand \@@startlink[1]{}%
\providecommand \@@endlink[0]{}%
\providecommand \url  [0]{\begingroup\@sanitize@url \@url }%
\providecommand \@url [1]{\endgroup\@href {#1}{\urlprefix }}%
\providecommand \urlprefix  [0]{URL }%
\providecommand \Eprint [0]{\href }%
\providecommand \doibase [0]{https://doi.org/}%
\providecommand \selectlanguage [0]{\@gobble}%
\providecommand \bibinfo  [0]{\@secondoftwo}%
\providecommand \bibfield  [0]{\@secondoftwo}%
\providecommand \translation [1]{[#1]}%
\providecommand \BibitemOpen [0]{}%
\providecommand \bibitemStop [0]{}%
\providecommand \bibitemNoStop [0]{.\EOS\space}%
\providecommand \EOS [0]{\spacefactor3000\relax}%
\providecommand \BibitemShut  [1]{\csname bibitem#1\endcsname}%
\let\auto@bib@innerbib\@empty
\bibitem [{\citenamefont {Kiger}\ and\ \citenamefont
  {Duncan}(2012)}]{kiger2012air}%
  \BibitemOpen
  \bibfield  {author} {\bibinfo {author} {\bibfnamefont {K.~T.}\ \bibnamefont
  {Kiger}}\ and\ \bibinfo {author} {\bibfnamefont {J.~H.}\ \bibnamefont
  {Duncan}},\ }\bibfield  {title} {\bibinfo {title} {Air-entrainment mechanisms
  in plunging jets and breaking waves},\ }\href@noop {} {\bibfield  {journal}
  {\bibinfo  {journal} {Annual Review of Fluid Mechanics}\ }\textbf {\bibinfo
  {volume} {44}},\ \bibinfo {pages} {563} (\bibinfo {year} {2012})}\BibitemShut
  {NoStop}%
\bibitem [{\citenamefont {Erinin}\ \emph {et~al.}(2023)\citenamefont {Erinin},
  \citenamefont {Liu}, \citenamefont {Liu}, \citenamefont {Mostert},
  \citenamefont {Deike},\ and\ \citenamefont {Duncan}}]{erinin2023effects}%
  \BibitemOpen
  \bibfield  {author} {\bibinfo {author} {\bibfnamefont {M.}~\bibnamefont
  {Erinin}}, \bibinfo {author} {\bibfnamefont {C.}~\bibnamefont {Liu}},
  \bibinfo {author} {\bibfnamefont {X.}~\bibnamefont {Liu}}, \bibinfo {author}
  {\bibfnamefont {W.}~\bibnamefont {Mostert}}, \bibinfo {author} {\bibfnamefont
  {L.}~\bibnamefont {Deike}},\ and\ \bibinfo {author} {\bibfnamefont
  {J.}~\bibnamefont {Duncan}},\ }\bibfield  {title} {\bibinfo {title} {The
  effects of surfactants on plunging breakers},\ }\href@noop {} {\bibfield
  {journal} {\bibinfo  {journal} {Journal of Fluid Mechanics}\ }\textbf
  {\bibinfo {volume} {972}},\ \bibinfo {pages} {R5} (\bibinfo {year}
  {2023})}\BibitemShut {NoStop}%
\bibitem [{\citenamefont {Liu}\ and\ \citenamefont
  {Duncan}(2003)}]{liu2003effects}%
  \BibitemOpen
  \bibfield  {author} {\bibinfo {author} {\bibfnamefont {X.}~\bibnamefont
  {Liu}}\ and\ \bibinfo {author} {\bibfnamefont {J.~H.}\ \bibnamefont
  {Duncan}},\ }\bibfield  {title} {\bibinfo {title} {The effects of surfactants
  on spilling breaking waves},\ }\href@noop {} {\bibfield  {journal} {\bibinfo
  {journal} {Nature}\ }\textbf {\bibinfo {volume} {421}},\ \bibinfo {pages}
  {520} (\bibinfo {year} {2003})}\BibitemShut {NoStop}%
\bibitem [{\citenamefont {Cheah}\ and\ \citenamefont
  {Cilliers}(2005)}]{cheah2005foaming}%
  \BibitemOpen
  \bibfield  {author} {\bibinfo {author} {\bibfnamefont {O.}~\bibnamefont
  {Cheah}}\ and\ \bibinfo {author} {\bibfnamefont {J.}~\bibnamefont
  {Cilliers}},\ }\bibfield  {title} {\bibinfo {title} {Foaming behaviour of
  aerosol ot solutions at low concentrations using a continuous plunging jet
  method},\ }\href@noop {} {\bibfield  {journal} {\bibinfo  {journal} {Colloids
  and Surfaces A: Physicochemical and Engineering Aspects}\ }\textbf {\bibinfo
  {volume} {263}},\ \bibinfo {pages} {347} (\bibinfo {year}
  {2005})}\BibitemShut {NoStop}%
\bibitem [{\citenamefont {Ross}\ and\ \citenamefont
  {Miles}(1941)}]{ross1941apparatus}%
  \BibitemOpen
  \bibfield  {author} {\bibinfo {author} {\bibfnamefont {J.}~\bibnamefont
  {Ross}}\ and\ \bibinfo {author} {\bibfnamefont {G.~D.}\ \bibnamefont
  {Miles}},\ }\bibfield  {title} {\bibinfo {title} {An apparatus for comparison
  of foaming properties of soaps and detergents},\ }\href@noop {} {\bibfield
  {journal} {\bibinfo  {journal} {Oil \& Soap}\ }\textbf {\bibinfo {volume}
  {18}},\ \bibinfo {pages} {99} (\bibinfo {year} {1941})}\BibitemShut {NoStop}%
\bibitem [{\citenamefont {Park}\ and\ \citenamefont
  {Durian}(1994)}]{park1994viscous}%
  \BibitemOpen
  \bibfield  {author} {\bibinfo {author} {\bibfnamefont {S.}~\bibnamefont
  {Park}}\ and\ \bibinfo {author} {\bibfnamefont {D.~J.}\ \bibnamefont
  {Durian}},\ }\bibfield  {title} {\bibinfo {title} {Viscous and elastic
  fingering instabilities in foam},\ }\href@noop {} {\bibfield  {journal}
  {\bibinfo  {journal} {Physical review letters}\ }\textbf {\bibinfo {volume}
  {72}},\ \bibinfo {pages} {3347} (\bibinfo {year} {1994})}\BibitemShut
  {NoStop}%
\bibitem [{\citenamefont {Mensire}\ \emph {et~al.}(2014)\citenamefont
  {Mensire}, \citenamefont {Piroird},\ and\ \citenamefont
  {Lorenceau}}]{mensire2014coalescence}%
  \BibitemOpen
  \bibfield  {author} {\bibinfo {author} {\bibfnamefont {R.}~\bibnamefont
  {Mensire}}, \bibinfo {author} {\bibfnamefont {K.}~\bibnamefont {Piroird}},\
  and\ \bibinfo {author} {\bibfnamefont {E.}~\bibnamefont {Lorenceau}},\
  }\bibfield  {title} {\bibinfo {title} {Coalescence of dry foam under water
  injection},\ }\href@noop {} {\bibfield  {journal} {\bibinfo  {journal} {Soft
  matter}\ }\textbf {\bibinfo {volume} {10}},\ \bibinfo {pages} {7068}
  (\bibinfo {year} {2014})}\BibitemShut {NoStop}%
\bibitem [{\citenamefont {Del~Prete}\ \emph {et~al.}(2013)\citenamefont
  {Del~Prete}, \citenamefont {Chinnayya}, \citenamefont {Domergue},
  \citenamefont {Hadjadj},\ and\ \citenamefont {Haas}}]{del2013blast}%
  \BibitemOpen
  \bibfield  {author} {\bibinfo {author} {\bibfnamefont {E.}~\bibnamefont
  {Del~Prete}}, \bibinfo {author} {\bibfnamefont {A.}~\bibnamefont
  {Chinnayya}}, \bibinfo {author} {\bibfnamefont {L.}~\bibnamefont {Domergue}},
  \bibinfo {author} {\bibfnamefont {A.}~\bibnamefont {Hadjadj}},\ and\ \bibinfo
  {author} {\bibfnamefont {J.-F.}\ \bibnamefont {Haas}},\ }\bibfield  {title}
  {\bibinfo {title} {Blast wave mitigation by dry aqueous foams},\ }\href@noop
  {} {\bibfield  {journal} {\bibinfo  {journal} {Shock waves}\ }\textbf
  {\bibinfo {volume} {23}},\ \bibinfo {pages} {39} (\bibinfo {year}
  {2013})}\BibitemShut {NoStop}%
\bibitem [{\citenamefont {Salem}\ \emph {et~al.}(2013)\citenamefont {Salem},
  \citenamefont {Cantat},\ and\ \citenamefont {Dollet}}]{salem2013response}%
  \BibitemOpen
  \bibfield  {author} {\bibinfo {author} {\bibfnamefont {I.~B.}\ \bibnamefont
  {Salem}}, \bibinfo {author} {\bibfnamefont {I.}~\bibnamefont {Cantat}},\ and\
  \bibinfo {author} {\bibfnamefont {B.}~\bibnamefont {Dollet}},\ }\bibfield
  {title} {\bibinfo {title} {Response of a two-dimensional liquid foam to air
  injection: Influence of surfactants, critical velocities and branched
  fracture},\ }\href@noop {} {\bibfield  {journal} {\bibinfo  {journal}
  {Colloids and Surfaces A: Physicochemical and Engineering Aspects}\ }\textbf
  {\bibinfo {volume} {438}},\ \bibinfo {pages} {41} (\bibinfo {year}
  {2013})}\BibitemShut {NoStop}%
\bibitem [{\citenamefont {Arif}\ \emph {et~al.}(2010)\citenamefont {Arif},
  \citenamefont {Tsai},\ and\ \citenamefont {Hilgenfeldt}}]{arif2010speed}%
  \BibitemOpen
  \bibfield  {author} {\bibinfo {author} {\bibfnamefont {S.}~\bibnamefont
  {Arif}}, \bibinfo {author} {\bibfnamefont {J.-C.}\ \bibnamefont {Tsai}},\
  and\ \bibinfo {author} {\bibfnamefont {S.}~\bibnamefont {Hilgenfeldt}},\
  }\bibfield  {title} {\bibinfo {title} {Speed of crack propagation in dry
  aqueous foam},\ }\href@noop {} {\bibfield  {journal} {\bibinfo  {journal}
  {Europhysics Letters}\ }\textbf {\bibinfo {volume} {92}},\ \bibinfo {pages}
  {38001} (\bibinfo {year} {2010})}\BibitemShut {NoStop}%
\bibitem [{\citenamefont {Salkin}\ \emph {et~al.}(2016)\citenamefont {Salkin},
  \citenamefont {Schmit}, \citenamefont {Panizza},\ and\ \citenamefont
  {Courbin}}]{salkin2016generating}%
  \BibitemOpen
  \bibfield  {author} {\bibinfo {author} {\bibfnamefont {L.}~\bibnamefont
  {Salkin}}, \bibinfo {author} {\bibfnamefont {A.}~\bibnamefont {Schmit}},
  \bibinfo {author} {\bibfnamefont {P.}~\bibnamefont {Panizza}},\ and\ \bibinfo
  {author} {\bibfnamefont {L.}~\bibnamefont {Courbin}},\ }\bibfield  {title}
  {\bibinfo {title} {Generating soap bubbles by blowing on soap films},\
  }\href@noop {} {\bibfield  {journal} {\bibinfo  {journal} {Physical review
  letters}\ }\textbf {\bibinfo {volume} {116}},\ \bibinfo {pages} {077801}
  (\bibinfo {year} {2016})}\BibitemShut {NoStop}%
\bibitem [{\citenamefont {Gilet}\ and\ \citenamefont
  {Bush}(2009)}]{gilet2009fluid}%
  \BibitemOpen
  \bibfield  {author} {\bibinfo {author} {\bibfnamefont {T.}~\bibnamefont
  {Gilet}}\ and\ \bibinfo {author} {\bibfnamefont {J.~W.}\ \bibnamefont
  {Bush}},\ }\bibfield  {title} {\bibinfo {title} {The fluid trampoline:
  droplets bouncing on a soap film},\ }\href@noop {} {\bibfield  {journal}
  {\bibinfo  {journal} {Journal of Fluid Mechanics}\ }\textbf {\bibinfo
  {volume} {625}},\ \bibinfo {pages} {167} (\bibinfo {year}
  {2009})}\BibitemShut {NoStop}%
\bibitem [{\citenamefont {Le~Goff}\ \emph {et~al.}(2014)\citenamefont
  {Le~Goff}, \citenamefont {Qu{\'e}r{\'e}},\ and\ \citenamefont
  {Clanet}}]{goff2014shooting}%
  \BibitemOpen
  \bibfield  {author} {\bibinfo {author} {\bibfnamefont {A.}~\bibnamefont
  {Le~Goff}}, \bibinfo {author} {\bibfnamefont {D.}~\bibnamefont
  {Qu{\'e}r{\'e}}},\ and\ \bibinfo {author} {\bibfnamefont {C.}~\bibnamefont
  {Clanet}},\ }\bibfield  {title} {\bibinfo {title} {Shooting in a foam},\
  }\href@noop {} {\bibfield  {journal} {\bibinfo  {journal} {Soft matter}\
  }\textbf {\bibinfo {volume} {10}},\ \bibinfo {pages} {6696} (\bibinfo {year}
  {2014})}\BibitemShut {NoStop}%
\bibitem [{\citenamefont {Le~Goff}\ \emph {et~al.}(2013)\citenamefont
  {Le~Goff}, \citenamefont {Cobelli},\ and\ \citenamefont
  {Lagubeau}}]{goff2013supershear}%
  \BibitemOpen
  \bibfield  {author} {\bibinfo {author} {\bibfnamefont {A.}~\bibnamefont
  {Le~Goff}}, \bibinfo {author} {\bibfnamefont {P.}~\bibnamefont {Cobelli}},\
  and\ \bibinfo {author} {\bibfnamefont {G.}~\bibnamefont {Lagubeau}},\
  }\bibfield  {title} {\bibinfo {title} {Supershear rayleigh waves at a soft
  interface},\ }\href@noop {} {\bibfield  {journal} {\bibinfo  {journal}
  {Physical Review Letters}\ }\textbf {\bibinfo {volume} {110}},\ \bibinfo
  {pages} {236101} (\bibinfo {year} {2013})}\BibitemShut {NoStop}%
\bibitem [{\citenamefont {Monloubou}\ \emph {et~al.}(2015)\citenamefont
  {Monloubou}, \citenamefont {Saint-Jalmes}, \citenamefont {Dollet},\ and\
  \citenamefont {Cantat}}]{monloubou2015influence}%
  \BibitemOpen
  \bibfield  {author} {\bibinfo {author} {\bibfnamefont {M.}~\bibnamefont
  {Monloubou}}, \bibinfo {author} {\bibfnamefont {A.}~\bibnamefont
  {Saint-Jalmes}}, \bibinfo {author} {\bibfnamefont {B.}~\bibnamefont
  {Dollet}},\ and\ \bibinfo {author} {\bibfnamefont {I.}~\bibnamefont
  {Cantat}},\ }\bibfield  {title} {\bibinfo {title} {Influence of bubble size
  and thermal dissipation on compressive wave attenuation in liquid foams},\
  }\href@noop {} {\bibfield  {journal} {\bibinfo  {journal} {Europhysics
  Letters}\ }\textbf {\bibinfo {volume} {112}},\ \bibinfo {pages} {34001}
  (\bibinfo {year} {2015})}\BibitemShut {NoStop}%
\bibitem [{\citenamefont {Monloubou}\ \emph {et~al.}(2016)\citenamefont
  {Monloubou}, \citenamefont {Bruning}, \citenamefont {Saint-Jalmes},
  \citenamefont {Dollet},\ and\ \citenamefont {Cantat}}]{monloubou2016blast}%
  \BibitemOpen
  \bibfield  {author} {\bibinfo {author} {\bibfnamefont {M.}~\bibnamefont
  {Monloubou}}, \bibinfo {author} {\bibfnamefont {M.~A.}\ \bibnamefont
  {Bruning}}, \bibinfo {author} {\bibfnamefont {A.}~\bibnamefont
  {Saint-Jalmes}}, \bibinfo {author} {\bibfnamefont {B.}~\bibnamefont
  {Dollet}},\ and\ \bibinfo {author} {\bibfnamefont {I.}~\bibnamefont
  {Cantat}},\ }\bibfield  {title} {\bibinfo {title} {Blast wave attenuation in
  liquid foams: role of gas and evidence of an optimal bubble size},\
  }\href@noop {} {\bibfield  {journal} {\bibinfo  {journal} {Soft matter}\
  }\textbf {\bibinfo {volume} {12}},\ \bibinfo {pages} {8015} (\bibinfo {year}
  {2016})}\BibitemShut {NoStop}%
\bibitem [{\citenamefont {Cohen}\ \emph {et~al.}(2017)\citenamefont {Cohen},
  \citenamefont {Fraysse},\ and\ \citenamefont {Raufaste}}]{cohen2017raufaste}%
  \BibitemOpen
  \bibfield  {author} {\bibinfo {author} {\bibfnamefont {A.}~\bibnamefont
  {Cohen}}, \bibinfo {author} {\bibfnamefont {N.}~\bibnamefont {Fraysse}},\
  and\ \bibinfo {author} {\bibfnamefont {C.}~\bibnamefont {Raufaste}},\
  }\bibfield  {title} {\bibinfo {title} {Melde’s experiment on a vibrating
  liquid foam microchannel},\ }\href@noop {} {\bibfield  {journal} {\bibinfo
  {journal} {Physical Review Letters}\ }\textbf {\bibinfo {volume} {119}},\
  \bibinfo {pages} {238001} (\bibinfo {year} {2017})}\BibitemShut {NoStop}%
\bibitem [{\citenamefont {Yanagisawa}\ \emph {et~al.}(2021)\citenamefont
  {Yanagisawa}, \citenamefont {Tani},\ and\ \citenamefont
  {Kurita}}]{tani2021dynamics}%
  \BibitemOpen
  \bibfield  {author} {\bibinfo {author} {\bibfnamefont {N.}~\bibnamefont
  {Yanagisawa}}, \bibinfo {author} {\bibfnamefont {M.}~\bibnamefont {Tani}},\
  and\ \bibinfo {author} {\bibfnamefont {R.}~\bibnamefont {Kurita}},\
  }\bibfield  {title} {\bibinfo {title} {Dynamics and mechanism of liquid film
  collapse in a foam},\ }\href@noop {} {\bibfield  {journal} {\bibinfo
  {journal} {Soft Matter}\ }\textbf {\bibinfo {volume} {17}},\ \bibinfo {pages}
  {1738} (\bibinfo {year} {2021})}\BibitemShut {NoStop}%
\bibitem [{\citenamefont {Kirstetter}\ \emph {et~al.}(2012)\citenamefont
  {Kirstetter}, \citenamefont {Raufaste},\ and\ \citenamefont
  {Celestini}}]{kirstetter2012}%
  \BibitemOpen
  \bibfield  {author} {\bibinfo {author} {\bibfnamefont {G.}~\bibnamefont
  {Kirstetter}}, \bibinfo {author} {\bibfnamefont {C.}~\bibnamefont
  {Raufaste}},\ and\ \bibinfo {author} {\bibfnamefont {F.}~\bibnamefont
  {Celestini}},\ }\bibfield  {title} {\bibinfo {title} {Jet impact on a soap
  film},\ }\href@noop {} {\bibfield  {journal} {\bibinfo  {journal} {Physical
  Review E—Statistical, Nonlinear, and Soft Matter Physics}\ }\textbf
  {\bibinfo {volume} {86}},\ \bibinfo {pages} {036303} (\bibinfo {year}
  {2012})}\BibitemShut {NoStop}%
\bibitem [{Note1()}]{Note1}%
  \BibitemOpen
  \bibinfo {note} {See video examples in the supplementary materials}\BibitemShut {NoStop}%
\bibitem [{\citenamefont {Gaichies}\ \emph {et~al.}(2024)\citenamefont
  {Gaichies}, \citenamefont {Salonen}, \citenamefont {Antkowiak},\ and\
  \citenamefont {Rio}}]{gaichies2024}%
  \BibitemOpen
  \bibfield  {author} {\bibinfo {author} {\bibfnamefont {T.}~\bibnamefont
  {Gaichies}}, \bibinfo {author} {\bibfnamefont {A.}~\bibnamefont {Salonen}},
  \bibinfo {author} {\bibfnamefont {A.}~\bibnamefont {Antkowiak}},\ and\
  \bibinfo {author} {\bibfnamefont {E.}~\bibnamefont {Rio}},\ }\bibfield
  {title} {\bibinfo {title} {Effective water/water contact angle at the base of
  an impinging jet},\ }\href@noop {} {\bibfield  {journal} {\bibinfo  {journal}
  {Physical Review Fluids}\ }\textbf {\bibinfo {volume} {9}},\ \bibinfo {pages}
  {034003} (\bibinfo {year} {2024})}\BibitemShut {NoStop}%
\bibitem [{\citenamefont {Van~der Walt}\ \emph {et~al.}(2014)\citenamefont
  {Van~der Walt}, \citenamefont {Sch{\"o}nberger}, \citenamefont
  {Nunez-Iglesias}, \citenamefont {Boulogne}, \citenamefont {Warner},
  \citenamefont {Yager}, \citenamefont {Gouillart},\ and\ \citenamefont
  {Yu}}]{van2014scikit}%
  \BibitemOpen
  \bibfield  {author} {\bibinfo {author} {\bibfnamefont {S.}~\bibnamefont
  {Van~der Walt}}, \bibinfo {author} {\bibfnamefont {J.~L.}\ \bibnamefont
  {Sch{\"o}nberger}}, \bibinfo {author} {\bibfnamefont {J.}~\bibnamefont
  {Nunez-Iglesias}}, \bibinfo {author} {\bibfnamefont {F.}~\bibnamefont
  {Boulogne}}, \bibinfo {author} {\bibfnamefont {J.~D.}\ \bibnamefont
  {Warner}}, \bibinfo {author} {\bibfnamefont {N.}~\bibnamefont {Yager}},
  \bibinfo {author} {\bibfnamefont {E.}~\bibnamefont {Gouillart}},\ and\
  \bibinfo {author} {\bibfnamefont {T.}~\bibnamefont {Yu}},\ }\bibfield
  {title} {\bibinfo {title} {scikit-image: image processing in python},\
  }\href@noop {} {\bibfield  {journal} {\bibinfo  {journal} {PeerJ}\ }\textbf
  {\bibinfo {volume} {2}},\ \bibinfo {pages} {e453} (\bibinfo {year}
  {2014})}\BibitemShut {NoStop}%
\bibitem [{\citenamefont {McGraw}\ \emph {et~al.}(2012)\citenamefont {McGraw},
  \citenamefont {Salez}, \citenamefont {B{\"a}umchen}, \citenamefont
  {Rapha{\"e}l},\ and\ \citenamefont {Dalnoki-Veress}}]{mcgraw2012self}%
  \BibitemOpen
  \bibfield  {author} {\bibinfo {author} {\bibfnamefont {J.~D.}\ \bibnamefont
  {McGraw}}, \bibinfo {author} {\bibfnamefont {T.}~\bibnamefont {Salez}},
  \bibinfo {author} {\bibfnamefont {O.}~\bibnamefont {B{\"a}umchen}}, \bibinfo
  {author} {\bibfnamefont {E.}~\bibnamefont {Rapha{\"e}l}},\ and\ \bibinfo
  {author} {\bibfnamefont {K.}~\bibnamefont {Dalnoki-Veress}},\ }\bibfield
  {title} {\bibinfo {title} {Self-similarity and energy dissipation in stepped
  polymer films},\ }\href@noop {} {\bibfield  {journal} {\bibinfo  {journal}
  {Physical Review Letters}\ }\textbf {\bibinfo {volume} {109}},\ \bibinfo
  {pages} {128303} (\bibinfo {year} {2012})}\BibitemShut {NoStop}%
\bibitem [{\citenamefont {Morton}\ \emph {et~al.}(1956)\citenamefont {Morton},
  \citenamefont {Taylor},\ and\ \citenamefont {Turner}}]{morton1956turbulent}%
  \BibitemOpen
  \bibfield  {author} {\bibinfo {author} {\bibfnamefont {B.~R.}\ \bibnamefont
  {Morton}}, \bibinfo {author} {\bibfnamefont {G.~I.}\ \bibnamefont {Taylor}},\
  and\ \bibinfo {author} {\bibfnamefont {J.~S.}\ \bibnamefont {Turner}},\
  }\bibfield  {title} {\bibinfo {title} {Turbulent gravitational convection
  from maintained and instantaneous sources},\ }\href@noop {} {\bibfield
  {journal} {\bibinfo  {journal} {Proceedings of the Royal Society of London.
  Series A. Mathematical and Physical Sciences}\ }\textbf {\bibinfo {volume}
  {234}},\ \bibinfo {pages} {1} (\bibinfo {year} {1956})}\BibitemShut {NoStop}%
\bibitem [{\citenamefont {Karol J.~Mysels}\ and\ \citenamefont
  {Frankel}(1959)}]{frankel}%
  \BibitemOpen
  \bibfield  {author} {\bibinfo {author} {\bibfnamefont {K.~S.}\ \bibnamefont
  {Karol J.~Mysels}}\ and\ \bibinfo {author} {\bibfnamefont {S.}~\bibnamefont
  {Frankel}},\ }\href@noop {} {\emph {\bibinfo {title} {"Soap films : studies
  of their thinning and a bibliography"}}}\ (\bibinfo  {publisher} {Pergamon
  Press},\ \bibinfo {year} {1959})\BibitemShut {NoStop}%
\bibitem [{sup()}]{supp}%
  \BibitemOpen
  \href@noop {} {}\bibinfo {note} {See Supplemental Material at
  URL-will-be-inserted-by-publisher for the data of the
  experiments.}\BibitemShut {Stop}%
\bibitem [{\citenamefont {Mysels}\ \emph {et~al.}(1961)\citenamefont {Mysels},
  \citenamefont {C},\ and\ \citenamefont {Skewis}}]{mysels61}%
  \BibitemOpen
  \bibfield  {author} {\bibinfo {author} {\bibfnamefont {K.~J.}\ \bibnamefont
  {Mysels}}, \bibinfo {author} {\bibfnamefont {M.~C.~M.}\ \bibnamefont {C}},\
  and\ \bibinfo {author} {\bibfnamefont {J.~D.}\ \bibnamefont {Skewis}},\
  }\bibfield  {title} {\bibinfo {title} {The measurement of film elasticity},\
  }\href@noop {} {\bibfield  {journal} {\bibinfo  {journal} {J. Phys. Chem.}\
  }\textbf {\bibinfo {volume} {65}},\ \bibinfo {pages} {1107} (\bibinfo {year}
  {1961})}\BibitemShut {NoStop}%
\bibitem [{\citenamefont {van~den Tempel}\ \emph {et~al.}(1965)\citenamefont
  {van~den Tempel}, \citenamefont {Lucassen},\ and\ \citenamefont
  {Lucassen-Reynders}}]{vandentempel65}%
  \BibitemOpen
  \bibfield  {author} {\bibinfo {author} {\bibfnamefont {M.}~\bibnamefont
  {van~den Tempel}}, \bibinfo {author} {\bibfnamefont {J.}~\bibnamefont
  {Lucassen}},\ and\ \bibinfo {author} {\bibfnamefont {E.~H.}\ \bibnamefont
  {Lucassen-Reynders}},\ }\bibfield  {title} {\bibinfo {title} {Application of
  surface thermodynamics to gibbs elasticity},\ }\href@noop {} {\bibfield
  {journal} {\bibinfo  {journal} {J. Phys. Chem.}\ }\textbf {\bibinfo {volume}
  {69}},\ \bibinfo {pages} {1798} (\bibinfo {year} {1965})}\BibitemShut
  {NoStop}%
\bibitem [{\citenamefont {Poryles}\ \emph {et~al.}(2022)\citenamefont
  {Poryles}, \citenamefont {Lenavetier}, \citenamefont {Schaub}, \citenamefont
  {Bussonni{\`e}re}, \citenamefont {Saint-Jalmes},\ and\ \citenamefont
  {Cantat}}]{poryles22}%
  \BibitemOpen
  \bibfield  {author} {\bibinfo {author} {\bibfnamefont {R.}~\bibnamefont
  {Poryles}}, \bibinfo {author} {\bibfnamefont {T.}~\bibnamefont {Lenavetier}},
  \bibinfo {author} {\bibfnamefont {E.}~\bibnamefont {Schaub}}, \bibinfo
  {author} {\bibfnamefont {A.}~\bibnamefont {Bussonni{\`e}re}}, \bibinfo
  {author} {\bibfnamefont {A.}~\bibnamefont {Saint-Jalmes}},\ and\ \bibinfo
  {author} {\bibfnamefont {I.}~\bibnamefont {Cantat}},\ }\bibfield  {title}
  {\bibinfo {title} {Non linear elasticity of foam films made of sds/dodecanol
  mixtures},\ }\href@noop {} {\ \textbf {\bibinfo {volume} {18}},\ \bibinfo
  {pages} {2046} (\bibinfo {year} {2022})}\BibitemShut {NoStop}%
\bibitem [{\citenamefont {Prins}\ \emph {et~al.}(1967)\citenamefont {Prins},
  \citenamefont {Arcuri},\ and\ \citenamefont {{van den Tempel}}}]{prins67}%
  \BibitemOpen
  \bibfield  {author} {\bibinfo {author} {\bibfnamefont {A.}~\bibnamefont
  {Prins}}, \bibinfo {author} {\bibfnamefont {C.}~\bibnamefont {Arcuri}},\ and\
  \bibinfo {author} {\bibfnamefont {M.}~\bibnamefont {{van den Tempel}}},\
  }\bibfield  {title} {\bibinfo {title} {Elasticity of thin liquid films},\
  }\href {https://doi.org/https://doi.org/10.1016/0021-9797(67)90281-0} {\
  \textbf {\bibinfo {volume} {24}},\ \bibinfo {pages} {84} (\bibinfo {year}
  {1967})}\BibitemShut {NoStop}%
\end{thebibliography}
%

\end{document}